\newif\ifsingle

\newif\iftemplate
\templatetrue 

\iftemplate
\documentclass{article}
\usepackage{amsmath,graphicx,spconf}
\else
\ifsingle
\documentclass[11pt,draftclsnofoot, onecolumn]{IEEEtran}		
\else		
\documentclass[9pt,final, twocolumn]{IEEEtran}
\fi
\fi

\usepackage{times}
\usepackage{amsmath,dsfont}
\usepackage{amssymb,amsthm}
\usepackage{epsfig,verbatim}
\usepackage{subfigure}
\usepackage{setspace}
\usepackage{color}
\usepackage{cite}
\usepackage{epstopdf}
\usepackage{graphics}
\usepackage{accents}
\usepackage{acronym}
\usepackage[bookmarks,colorlinks]{hyperref}
\usepackage{booktabs}
\usepackage{mathtools}
\usepackage{float}
\usepackage[linesnumbered,ruled,lined]{algorithm2e}
\usepackage{algpseudocode}
\usepackage{enumitem}
\makeatletter
\newcommand{\removelatexerror}{\let\@latex@error\@gobble}
\makeatother

\def\BibTeX{{\rm B\kern-.05em{\sc i\kern-.025em b}\kern-.08em
    T\kern-.1667em\lower.7ex\hbox{E}\kern-.125emX}}
 \usepackage{setspace}
\makeatletter
\newcommand{\oset}[3][0ex]{%
  \mathrel{\mathop{#3}\limits^{
    \vbox to#1{\kern-2\ex@
    \hbox{$\scriptstyle#2$}\vss}}}}
\makeatother

\newcounter{relctr} 
\everydisplay\expandafter{\the\everydisplay\setcounter{relctr}{0}} 

\newcommand{\algalign}[2]
{\makebox[\maxwidth][r]{$#1{}$}${}#2$}

\AtBeginDocument{} 

\newcommand{\myVec}[1]{{\boldsymbol{#1}}}

\newcommand{\mySet}[1]{\mathcal{#1}}
	
\newcommand{\Ntraining}{T}

\acrodef{dnn}[DNN]{Deep Neural Network} 
\acrodef{csi}[CSI]{channel state information}
\acrodef{map}[MAP]{maximum a-posteriori probability}
\acrodef{snr}[SNR]{Signal-to-Noise Ratio}
\acrodef{ser}[SER]{symbol error rate}
\acrodef{bs}[BS]{Base Station} 
\acrodef{ml}[ML]{machine learning} 
\acrodef{iot}[IOT]{Interent of Things}
\acrodef{mimo}[MIMO]{multiple-input multiple-output}
\acrodef{pdf}[PDF]{probability density function}
\acrodef{rv}[RV]{random variable}
\acrodef{em}[EM]{electromagnetic}
\acrodef{hmm}[HMM]{hidden Markov model}
\acrodef{pdf}[PDF]{probability density function}
\acrodef{isi}[ISI]{intersymbol interference}   
\acrodef{fec}[FEC]{forward error correction}
\acrodef{sova}[SOVA]{soft-output Viterbi algorithm}
\acrodef{adc}[ADC]{analog-to-digital convertor}
\acrodef{cs}[CS]{compressed sensing}
\acrodef{dtft}[DTFT]{discrete-time Fourier transform}
\acrodef{csi}[CSI]{channel state information}
\acrodef{bpsk}[BPSK]{binary phase shift keying}
\acrodef{map}[MAP]{maximum a-posteriori probability}
\acrodef{bs}[BS]{Base Station} 
\acrodef{iot}[IOT]{Interent of Things}
\acrodef{mimo}[MIMO]{multiple-input multiple-output}
\acrodef{pdf}[PDF]{probability density function}
\acrodef{rv}[RV]{random variable}
\acrodef{ml}[ML]{machine learning}
\acrodef{fec}[FEC]{forward error correction}
\acrodef{rs}[RS]{Reed-Solomon}
\acrodef{lti}[LTI]{linear time-invariant}
\acrodef{wss}[WSS]{wide-sense stationary}
\acrodef{psd}[PSD]{power spectral density}
\acrodef{ser}[SER]{symbol error rate} 
\acrodef{ber}[BER]{bit error rate} 
\acrodef{gd}[GD]{gradient descent}
\acrodef{sgd}[SGD]{stochastic gradient descent} 
\acrodef{isi}[ISI]{intersymbol interference}  
\acrodef{awgn}[AWGN]{additive white Gaussian noise} 
\acrodef{ut}[UT]{user terminal} 
\acrodef{mmw}[mmWave]{millimeter wave}
\acrodef{noma}[NOMA]{non-orthognal multiple access}
\acrodef{mac}[MAC]{mulitple access channel}
\acrodef{fl}[FL]{Federated learning}
\acrodef{lstm}[LSTM]{long short-term memory}
\acrodef{ris}[RIS]{Reconfigurable Intelligent Surface}

\acrodef{ga}[GA]{Genetic Algorithm}
\acrodef{relu}[ReLU]{Rectified Linear Unit}
\acrodef{rx}[RX]{Receiver}
\acrodef{mse}[MSE]{Mean Squared Error}

\definecolor{blue}{rgb}{0,0,1}

\ifsingle

\else

\fi 

\iftemplate 

\title{Deep-Learning-Assisted Configuration of Reconfigurable Intelligent Surfaces in Dynamic rich-scattering Environments}

\name{Kyriakos Stylianopoulos, Nir Shlezinger, Philipp del Hougne, and George C. Alexandropoulos
\thanks{
	 K. Stylianopoulos and G.~C.~Alexandropoulos are with the Department of Informatics and Telecommunications, National and Kapodistrian University of Athens, 15784 Athens, Greece (e-mail:  \{kstylianop; alexandg\}@di.uoa.gr). 
	N. Shlezinger is with the School of ECE, Ben-Gurion University of the Negev, Beer-Sheva, Israel (e-mail: nirshl@bgu.ac.il). 
    P. del Hougne is with Univ Rennes, CNRS, IETR - UMR 6164, F-35000, Rennes, France (e-mail: philipp.del-hougne@univ-rennes1.fr). The work was supported by the EU H2020 RISE-6G project under grant number 101017011.
}}

\address{\vspace{-25mm}}
\else
\IEEEoverridecommandlockouts

\title{Deep Learning Configuration of Reconfigurable Intelligent Surfaces for rich-scattering Environments
}
\author{
	\IEEEauthorblockN{Kyriakos Stylianopoulos, Nir Shlezinger, Philipp del Hougne, and George C. Alexandropoulos 
	} 
	\thanks{K. Stylianopoulos and G.~C.~Alexandropoulos are with the Department of Informatics and Telecommunications, National and Kapodistrian University of Athens, 15784 Athens, Greece (e-mail:  \{kstylianop; alexandg\}@di.uoa.gr). 
	N. Shlezinger is with the School of ECE, Ben-Gurion University of the Negev, Beer-Sheva, Israel (e-mail: nirshl@bgu.ac.il). 
    P. del Hougne is with Univ Rennes, CNRS, IETR - UMR 6164, F-35000, Rennes, France (e-mail: philipp.del-hougne@univ-rennes1.fr).}

	\vspace{-0.5cm}
	
}
\vspace{-0.75cm}
\fi

\begin{document}

\maketitle
 \pagestyle{empty}
 \thispagestyle{empty}
 
\begin{abstract}
	
	The integration of \acp{ris} into wireless environments endows channels with programmability, and is  expected to play a key role in future communication standards. To date, most RIS-related efforts focus on quasi-free-space, where wireless channels are typically modeled analytically. Many realistic communication scenarios occur, however, in rich-scattering environments which, moreover, evolve dynamically. These conditions present a tremendous challenge in identifying an RIS configuration that optimizes the achievable communication rate. In this paper, we make a first step toward tackling this challenge. Based on a simulator that is faithful to the underlying wave physics, we train a deep neural network as surrogate forward model to capture the stochastic dependence of wireless channels on the RIS configuration under dynamic rich-scattering conditions. Subsequently, we use this model in combination with a genetic algorithm to identify RIS configurations optimizing the communication rate. We numerically demonstrate the ability of the proposed approach to tune \acp{ris} to improve the achievable rate in rich-scattering setups.

{\textbf{\textit{Index terms---}} Reconfigurable intelligent surfaces, deep learning, rich-scattering, dynamic wireless environments.}  
\end{abstract}

\acresetall 

\vspace{-0.2cm}
\section{Introduction}
\vspace{-0.1cm}
	
	Recent years have witnessed the emergence of programmable wireless environments, enabled by using \acp{ris}, as a disruptive new wireless networking paradigm~\cite{subrt2012intelligent,huang2019reconfigurable,di2019smart,alexandropoulos2021reconfigurable,RISE6G_COMMAG}. Programmable wireless channels open up a host of new opportunities in wireless communications and sensing. The majority of RIS-based ideas has to date been explored under the assumption of quasi-free-space, possibly with a few known scatterers. These conditions enable the deployment of analytical channel models for wirelesss propagation. In practice, estimating the channels in \ac{ris}-aided communication is expected to be complex and costly \cite{wang2020channel, liu2019matrix, hu2019two, alexandropoulos2021hybrid, LZAWFM2020}. 
	
	The associated challenges become more complex when we turn to rich-scattering environments. 
	Indoor environments inside buildings, metro stations, and vessels or airplanes often act as irregularly shaped scattering enclosures that give rise to significant reverberation. Wave propagation under these rich-scattering conditions strongly differs from the intensely studied free-space case \cite{alexandropoulos2021reconfigurable}. Previous  works explored the optimization of communication-related metrics in RIS-enabled \textit{static} rich-scattering enclosures~\cite{del2016spatiotemporal,del2019optimally,alexandropoulos2021reconfigurable}, focusing on enforcing pulse-like channel impulse responses for simple modulation scenarios based on either iterative experimental optimization of the \ac{ris}~\cite{del2016spatiotemporal,del2019optimally,alexandropoulos2021reconfigurable,zhou2021modeling}, or on the availability of channel state information \cite{arslan2021over,zhang2021spatial}. 
	However, realistic rich-scattering environments are rarely static~\cite{del2020robust}. The motion of inhabitants, rotating fans, and other factors yield a \textit{dynamic} nature of such environments that results in fast fading of the wireless channels, impacting communication design and limiting the ability to rely on channel knowledge. 
	
	In this paper, we make a first step toward tackling the challenge of identifying an RIS configuration that optimizes the communication rate in a dynamic rich-scattering environment. Given the overwhelming complexity of the rich-scattering environment, an analytical explicit treatment is intractable. Instead, our approach embraces the stochastic nature of the channel coefficients under such conditions. We capture the link between the RIS configuration and the key statistical parameter of the channel that determines the communication rate using a \ac{dnn}. Specifically, we use a low-\ac{snr} approximation of the ergodic achievable rate in rich-scattering wireless channels, which can be computed from their statistical moments, i.e., using the output of the trained \ac{dnn}. Based on this learned DNN model, we apply a genetic algorithm to optimize the RIS configuration in light of the ergodic capacity objective, inspired by the success of similar techniques in the context of deep priors \cite{bora2017compressed,shlezinger2020model}.  We numerically  demonstrate that the proposed \ac{dnn}-assisted optimization tunes the \ac{ris} to support rates within a minor gap of the maximal achievable values, without relying on explicit knowledge of the channel statistics and its interplay with the \ac{ris} setting. 
	
\vspace{-0.2cm}
\section{System Model}
\label{sec:Model}

\vspace{-0.2cm}
\subsection{Channel Model}
\label{subsec:Channel}
\vspace{-0.1cm} 
We consider a rich-scattering scenario in which a transmitter is communicating with a receiver inside an irregularly-shaped metallic enclosure that is equipped with an \ac{ris}. The channels exhibit rich-scattering due to reverberation inside the complex scattering enclosure resulting in multiple reflections off the walls. A perturbing object rotates in an uncontrolled manner inside the environment, yielding the dynamic nature of the fading wireless channel~\cite{del2020robust}. The \ac{ris} is comprised of $N$ reflecting elements, where each element can take values in a set $\mySet{S}$ representing the possible states one can configure. Since \acp{ris} typically admit a finite number of configurations, e.g., when controlled via PIN diodes \cite{del2016spatiotemporal,del2019optimally,dai2020reconfigurable}, we assume that $\mySet{S}$ is finite (specifically, binary), and denote the overall  configuration of the \ac{ris} by the vector $\myVec{\varphi}\in\mySet{S}^N$. 

The channel perturbations are assumed to be sufficiently fast to interpret the input-output relationship as a fast-fading frequency-selective channel in discrete time.
Let $h(t)$ be the stochastic channel impulse response,  encapsulating both rays traversing via the \ac{ris} as well as those which do not encounter it. While the statistics of $h(t)$ are affected by the \ac{ris} setting $\myVec{\varphi}$, this dependence may be extremely complex and even intractable \cite{alexandropoulos2021reconfigurable}. Letting $y(t)$ be the channel output observed by the receiver, the input-output relationship can be expressed as:
\begin{equation}
\label{eqn:ChannelModel}
    y(t) = h(t) \ast x(t) + w(t), \quad t\in\mathbb{Z},
\end{equation}
where $ \ast$ is the convolution operator, $x(t)$ is the channel input, and $w(t)$ is the white Gaussian noise with variance $\sigma^2$. We assume that the transmission bandwidth is divided into $B$ bins, not smaller than the coherence bandwidth, with central frequencies $f_1,f_2,\ldots, f_B$, and use $\{\rho(f_i)\}_{i=1}^{B}$ to denote the input spectral power allocation. Our DNN approach's training data is obtained from a simulator that faithfully represents the wave physics of the underlying rich-scattering scenario \cite{del2020robust}; the channel model and simulator are detailed in~\cite{PhysFad}.  

\vspace{-0.27cm}
\subsection{Problem Formulation}
\label{subsec:Problem}
\vspace{-0.1cm}  
 Our goal is to design a scheme for tuning the \ac{ris} configuration $\myVec{\varphi}$ for a given input spectral power allocation $\{\rho(f_i)\}_{i=1}^{B}$ in order to maximize the rate of achievable reliable communications.
 As the channel is fast-fading, we aim to maximize the ergodic achievable rate, given by \cite[Ch. 4.3]{goldsmith2005wireless} as
\begin{equation}
\label{eqn:ErgodicCap}
    R(\myVec{\varphi}) \!\triangleq\!  \mathbb{E}_{{\{H(f_i)\}}_{i=1}^B} \left\{\sum_{i=1}^{B} \log_2\! \left(1\!+\!\frac{|H(f_i)|^2\rho(f_i)}{\sigma^2} \right) \right\}\!,
\end{equation}
where $H(f) \triangleq\mathcal{F}\{h(t)\}$ is the stochastic channel frequency response between the single-antenna transmitter and receiver at frequency bin $f$, whose statistics depend on $\myVec{\varphi}$, while $\mathcal{F}(\cdot)$ is the discrete Fourier transform with $B$ bins. 
While the explicit relationship between the channel frequency response and a specific \ac{ris} configuration is unknown, extremely complex, and dependent on the perturber orientation, we have access to a training set of channels and their corresponding configuration, i.e., to $\Ntraining$ pairs of the form $\{\myVec{\varphi}_t,  \{H_t(f_i)\}_{i=1}^B \}_{t=1}^{\Ntraining}$.

\vspace{-0.2cm}
\section{RIS Configuration Methods}
\label{sec:Setting}
\vspace{-0.1cm}

\vspace{-0.2cm}
\subsection{Rationale}
\label{subsec:Rationale}
\vspace{-0.1cm}      
Determining the \ac{ris} configuration which maximizes the ergodic rate in \eqref{eqn:ErgodicCap} is associated with two core challenges: $1)$ computing the stochastic expectation with respect to the distribution of $\{H(f_i)\}_{i=1}^B$; and $2)$ identifying the relationship between the \ac{ris} configuration $\myVec{\varphi}$ and the distribution of $\{H(f_i)\}_{i=1}^B$. Neither of these tasks appears to be solvable with an analytical model of the wireless channels. 

To address the first challenge, we use a low-\ac{snr} approximation of the ergodic rate as our RIS-optimization objective:
\begin{equation}
\label{eqn:ErgodicCap2}
    \tilde{R}(\myVec{\varphi}) \triangleq  \sum_{i=1}^{B} \log_2 \left(1+\frac{\mathbb{E}\{|H(f_i)|^2\}\rho(f_i)}{\sigma^2} \right).
\end{equation}
The objective \eqref{eqn:ErgodicCap2}  only requires a characterization of the relationship between $\myVec{\varphi}$ and the second-order moments of the channel frequency responses. To tackle the second challenge, we train a dedicated \ac{dnn} which maps $\myVec{\varphi}$ to the second-order moments of the channel frequency responses. Ultimately, we will utilize this DNN to find the RIS configuration which optimizes \eqref{eqn:ErgodicCap2} using \acp{ga}, as detailed next.

\vspace{-0.2cm}
\subsection{Optimizing RIS Configuration}
\label{subsec:method}
\vspace{-0.1cm}   
Our proposed method for optimizing the \ac{ris} configuration based on the objective \eqref{eqn:ErgodicCap2} consists of two components: A \ac{dnn} which learns to capture the relationship between $\myVec{\varphi}$ and $\myVec{m} \triangleq \big[\mathbb{E}\{|H(f_1)|^2\}, \mathbb{E}\{|H(f_2)|^2\}, \ldots, \mathbb{E}\{|H(f_B)|^2\}\big]$; and a GA using the \ac{dnn} for optimizing the \ac{ris} configuration.

{\bf \ac{ris}-Channel \ac{dnn}:}
In order to learn how to map $\myVec{\varphi}$ into an estimate of $\myVec{m}$, we use a regression \ac{dnn} whose exact architecture is described in Section \ref{sec:Sims}.
In order to learn the parameters of the DNN, denoted by $\myVec{\theta}$, we first cluster the available data $\{\myVec{\varphi}_t,  \{H_t(f_i)\}_{i=1}^B \}_{t=1}^{\Ntraining}$ into $C\leq |\mySet{S}|^N$ clusters based on the \ac{ris} setting $\myVec{\varphi}_t$. For each cluster of index $c=1,2,\ldots,C$, we estimate $\{\mathbb{E}\{|H(f_i)|^2\}\}_{i=1}^B$ by averaging over the squared magnitude of the corresponding measured channels into the vector $\hat{\myVec{m}}_c\in \mySet{R}_{+}^B$. The resulting dataset, denoted by $\mySet{D}\triangleq\{\myVec{\varphi}_c, \hat{\myVec{m}}_c \}_{c=1}^{C}$, is used for training the \ac{dnn} based on the \ac{mse} loss function with an added  $\ell_2$-regularization term to avoid overfitting, i.e.:
\begin{equation}
\label{eqn:TrainingLoss}
    \mySet{L}(\myVec{\theta}) = \sum_{c=1}^{C}\|\hat{\myVec{m}}_{\myVec{\theta}}(\myVec{\varphi}_c) - \hat{\myVec{m}}_c\|^2 + \lambda \| \myVec{\theta} \|^2.\vspace{-0.1cm}
\end{equation}
In \eqref{eqn:TrainingLoss}, $\hat{\myVec{m}}_{\myVec{\theta}}(\cdot)$ denotes the \ac{dnn} mapping with parameters $\myVec{\theta}$, and $\lambda$ is the regularization factor.

{\bf Optimizing $\myVec{\varphi}$:}
Once the DNN is trained,  we use it to determine the setting $\myVec{\varphi}$ that maximizes \eqref{eqn:ErgodicCap2} by approximating $\{\mathbb{E}\{|H(f_i)|^2\}\}_{i=1}^B$ with the network's prediction $\hat{\myVec{m}}_{\myVec{\theta}}(\cdot)$. 
Since $\myVec{\varphi}$ takes values in the discrete set $\mySet{S}^N$, we do so using discrete optimization.
We employ a \ac{ga}, which is described in the sequel for binary configurations (i.e., $|\mySet{S}|=2$), corresponding to the \ac{ris} considered in Section \ref{sec:Sims} in line with current experimental prototypes~ \cite{del2016spatiotemporal,del2019optimally,alexandropoulos2021reconfigurable,dai2020reconfigurable,zhou2021modeling}. However, the algorithm can be adapted to any finite $|\mySet{S}|$ in the same manner, since \ac{ga} is applicable to arbitrary discrete optimization \cite{MitchellGeneticAlgorithmsBook}.

At every iteration \textit{(generation)} of index $k$, the \ac{ga} maintains a set \textit{(population)} $\mathcal{P}^{(k)}$ of size $l_{\rm pop}$ with candidate \ac{ris} configurations \textit{(solutions)}, which are used for evaluating the objective \textit{(fitness)} function, i.e.,~\eqref{eqn:ErgodicCap2}.
The next generation is then determined by the current population's \textit{offspring}, which is produced by applying the following steps in sequence:\\
    {\em 1. Tournament selection:} A multi-set, with length $l_{\rm pop}$, of \textit{parent} solutions is created by repeating the following process for each element: $k_{\rm tour}$ solutions from the current generation are sampled with replacement to participate. The solution with the highest fitness score is selected as a parent.\\
     {\em 2. Uniform crossover:} 
     The crossover mechanism involves creating two new candidate solution-vectors \textit{(offspring)} by combining information from candidates of the past generation \textit{(parents)}. Parents are assigned to pairs randomly. For each pair, two offspring solutions are generated. The first is constructed by selecting each bit from either parent with equal probability, and the latter as its ones-complement.\\
     {\em 3. Flip-bit mutation:} Each offspring solution has each of its bits flipped with probability $p_{\rm mut}$.

The above procedure results in an updated generation of equal size $l_{\rm pop}$.
The optimization is accomplished due to the fact that, solutions with high fitness scores participate in the evolutionary process to exchange genetic information, while the imposed stochasticity in all three steps leads to efficient exploration of the search space.
As a standard practice, the solution with the maximum fitness score discovered at any generation is kept at the end.
The resulting Deep RIS Setting algorithm steps are summarized in Algorithm~\ref{alg:deep-RIS-setting-genetic}.

\begin{algorithm}[!t]
	\caption{Proposed Deep RIS Setting}
	\label{alg:deep-RIS-setting-genetic}
	\KwData{Dataset $\mySet{D}=\{\myVec{\varphi}_c, \hat{\myVec{m}}_c \}_{c=1}^{C}$, $k_{\max}$, and $l_{\rm pop}$.}
	Train \ac{dnn} using $\mySet{D}$ based on loss \eqref{eqn:TrainingLoss} to obtain  $\myVec{\theta}$.\\
	Randomly initialize population $\mySet{P}^{(1)} = \{\myVec{\varphi}^{(1)}_j\}_{j=1}^{l_{\rm pop}}$.\\
	Set $s_{\rm max} = - \infty$ and $\myVec{\varphi}_{\rm max} = \emptyset$.\\
	\For{$k=1,2,\ldots,k_{\max}$}{
	    \For{$j=1,2,\ldots,l_{\rm pop}$}{
	        Calculate $\tilde{R}(\myVec{\varphi}^{(k)}_j)$ from \eqref{eqn:ErgodicCap2} using $\hat{\myVec{m}}_{\myVec{\theta}}(\myVec{\varphi}^{(k)}_j)$. \\
	        Set score $s^{(k)}_j = \tilde{R}(\myVec{\varphi}^{(k)}_j)$. \\
	        \If{$s^{(k)}_j > s_{\rm max}$}{
	            Set $s_{\rm max} = s^{(k)}_j$ and $\myVec{\varphi}_{\rm max} = \myVec{\varphi}^{(k)}_j$.\\
	        }
	    }
	    Generate  $\mySet{P}^{(k+1)}$ from $\mySet{P}^{(k)}$ using $\{s_j^{(k)}\}_{j=1}^{l_{\rm pop}}$ .\\
	}
	\KwOut{\ac{ris} configuration $\myVec{\varphi}_{\rm max}$.}
\end{algorithm}

\vspace{-0.2cm}
\subsection{Discussion}
\label{subsec:discussion}
\vspace{-0.1cm}      
Algorithm \ref{alg:deep-RIS-setting-genetic} enables the optimization of the \ac{ris} configuration under fast-fading frequency selective channels, such as those arising in rich-scattering conditions. It bypasses the need to impose a model on the relationship between the \ac{ris} configuration and the channel frequency response, in a manner amenable to discrete optimization using deep learning techniques. As opposed to  \cite{alexandropoulos2021reconfigurable}, where \acp{dnn} were trained to capture the instantaneous channel realization, here we account for the inherent stochasticity of rich-scattering channels and design our \ac{dnn} to estimate their statistical moments. Note that the \ac{dnn} does not try to learn the ergodic rate directly, even though it is the metric we ultimately try to optimize. This was decided because rate calculations are dependent on noise and spectral allocation values, hence, having a \ac{dnn} estimating the ergodic rate would imply that it would have to be retrained multiple times in setups of varying characteristics. 

Our current formulation considers a scalar point-to-point channel, and does not account for possible side information arising from, e.g., knowledge of the location of the communicating devices. When such additional knowledge is present, one can potentially incorporate it into the \ac{ris}-channel \ac{dnn} as a form of a hypernetwork \cite{ha2016hypernetworks}. Furthermore, Algorithm~\ref{alg:deep-RIS-setting-genetic} can also be extended to multi-user and multi-antenna systems, by, e.g., replacing the ergodic rate objective with the ergodic sum-rate for either uplink or downlink systems. Finally, our derivation of Algorithm~\ref{alg:deep-RIS-setting-genetic} considers a fixed input  power allocation $\rho(f)$. One can utilize the proposed approach to also optimize $\rho(f)$ in a joint manner with the \ac{ris} configuration. We leave the aforementioned extensions for future work.


\vspace{-0.2cm}
\section{Numerical Results}
\label{sec:Sims}
\vspace{-0.1cm}

We now apply our proposed method in a numerial study\footnote{
    The paper's TensorFlow code and used data are available at: \url{https://github.com/NoesysLab/Deep_RIS_Tuning_for_Rich_Scattering_Environments}.}. To that aim, we have simulated a rich-scattering environment with a single-antenna access point, a fixed-position and single-antenna \ac{rx}, and an \ac{ris} consisting of $N=21$ binary-tunable elements. We selected on purpose a relatively ``small'' number of possible RIS configurations (i.e., $2^{21}$), because they can be scanned with an exhaustive search, so that we can identify the globally optimal RIS configuration for rate maximization. 
To generate the rich-scattering environment, we have developed a simulator based on ~\cite{del2020robust,PhysFad}, simulating a rotating object that perturbs the environment.
The transmission observed at the \ac{rx} has been evaluated over $B=30$ frequency bins.

The data preparation step involves attaining mean frequency response measurements for different \ac{ris} profiles, as described in Section~\ref{subsec:method}. At first, $C=500$ \ac{ris} configurations, $\{\myVec{\varphi}_c\}_{c=1}^{C}$, were randomly selected. For each $\myVec{\varphi}_c$, we have set the perturber orientation to $100$ random angles, thus generating stochastic $\{H(f_i)\}_{i=1}^B$ measurements, which were then averaged out (per frequency point) to produce the target vectors $\myVec{\hat{m}}_c$.
Finally, $\mySet{D}$ was constructed as $\{\myVec{\varphi}_c, \hat{\myVec{m}}_c \}_{c=1}^{C}$ and we allocated $80\%$ of the data to the training set, whereas the remaining set was split in half to validation and test subsets.

\begin{figure}[tb]
     \centering
     \includegraphics[width=0.4\textwidth]{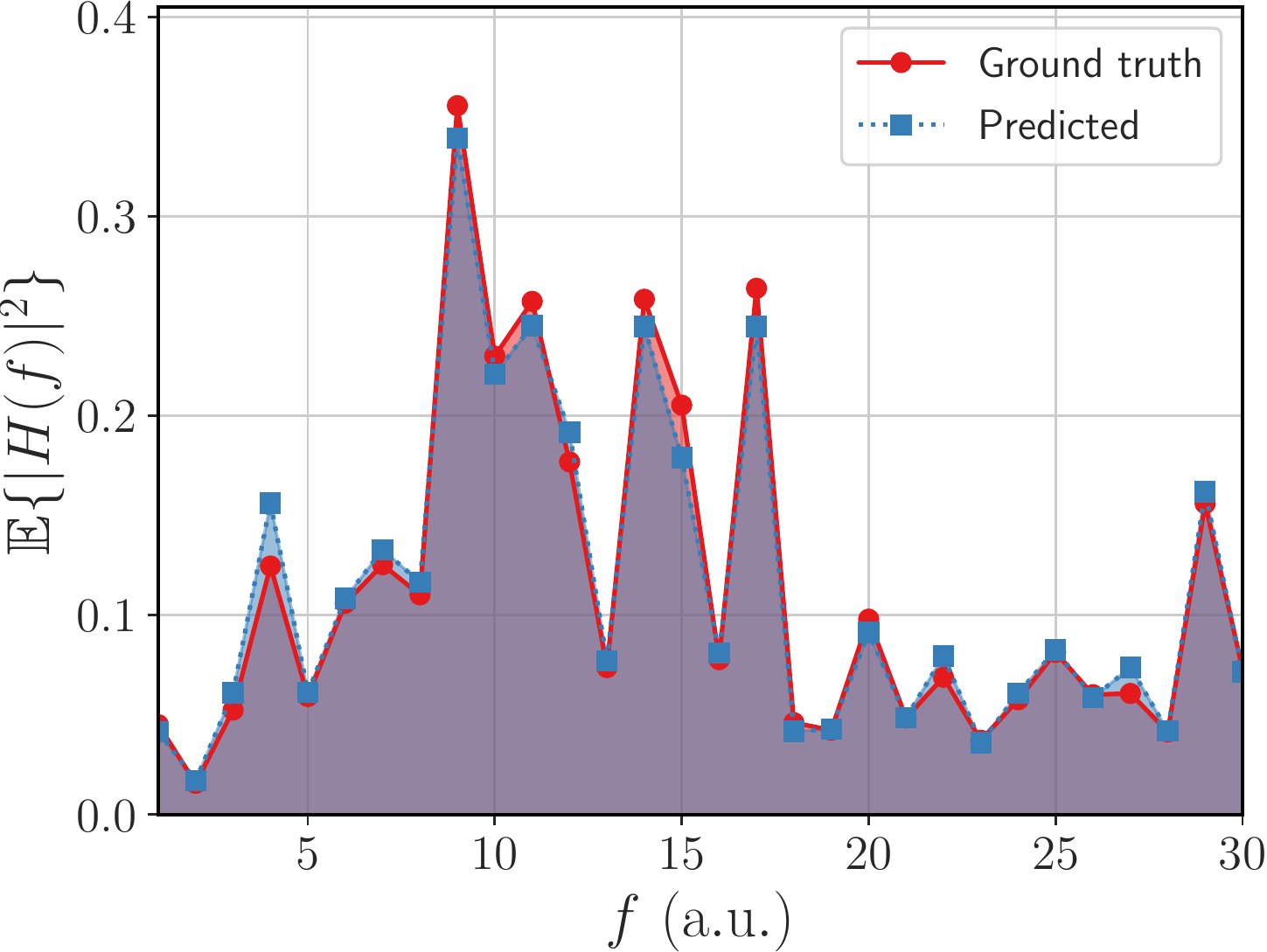}
     \vspace{-0.3cm}
    \caption{Ground truth and predicted mean frequency response of a random \ac{ris} configuration on which the network has not been trained. The frequency points are in arbitrary units (a.u.).}
    \label{fig:frequency_response_prediction}
\end{figure}

The \ac{ris}-channel \ac{dnn} is comprised of two fully connected hidden layers of $64$ units, each with ReLU activations. The output layer has $B$ neurons with linear activation to facilitate the regression process. Since the size of the data set is relatively small, we have set the regularization factor to $\lambda = 2.5 \times 10^{-5}$, and trained using Adam \cite{Kingma2015AdamAM}  with a learning rate of $0.001$. The hyper-parameter values for the learning rate, number of layers, units, and regularization factor were selected through grid search over possible combinations as the ones that minimize the \ac{mse} of the validation set.
The model was trained using early stopping, based on the validation \ac{mse}, for a total of $883$ epochs. The resulting \ac{mse} of the trained model on the test set was $1.27\times 10^{-4}$, which signifies that the network was indeed capable of learning channel statistics even in rich-scattering environments.
To illustrate the \ac{dnn} accuracy, the true and predicted squared magnitudes over the $B=30$ bins for a random \ac{ris} profile are illustrated in Fig.~\ref{fig:frequency_response_prediction}. Clearly, the trained \ac{dnn} succeeds in capturing the second-order moments of the channel frequency response.

Having trained and evaluated the \ac{ris}-channel \ac{dnn}, we proceed to apply Algorithm~\ref{alg:deep-RIS-setting-genetic} in order to identify a $\myVec{\varphi}$ that maximizes the ergodic rate in \eqref{eqn:ErgodicCap2}. Apart from Algorithm~\ref{alg:deep-RIS-setting-genetic}, we consider the baseline of selecting the best  out of $5000$ randomly evaluated candidates (Top random). Furthermore, a lower and an upper bound are reported, which are constructed by taking the average rate out of the $5000$ random configurations (Average) and by exhaustively evaluating all \ac{ris} configurations (Exhaustive). 
By setting the power allocation $\rho(f)$ to unity at all frequencies, we compare the achievable rates of the above methods versus the \ac{snr}. 
To obtain appropriate hyper-parameter values for our \ac{ga}, a short grid search was implemented for determining the values of $l_{\rm pop}$, $k_{\rm tour}$, and $p_{\rm mut}$. Since the action space was finite, allowing for a vast examination would be amenable to exhaustive search. Therefore, we allocated $30000$ 
evaluations to search between $60$ possible parameter combinations. The ones with the best performance were kept and the final algorithm was executed for $20000$ evaluations (i.e., $20000/l_{\rm pop}$ generations).

\begin{figure}[tb]
     \centering
     \includegraphics[width=0.4\textwidth]{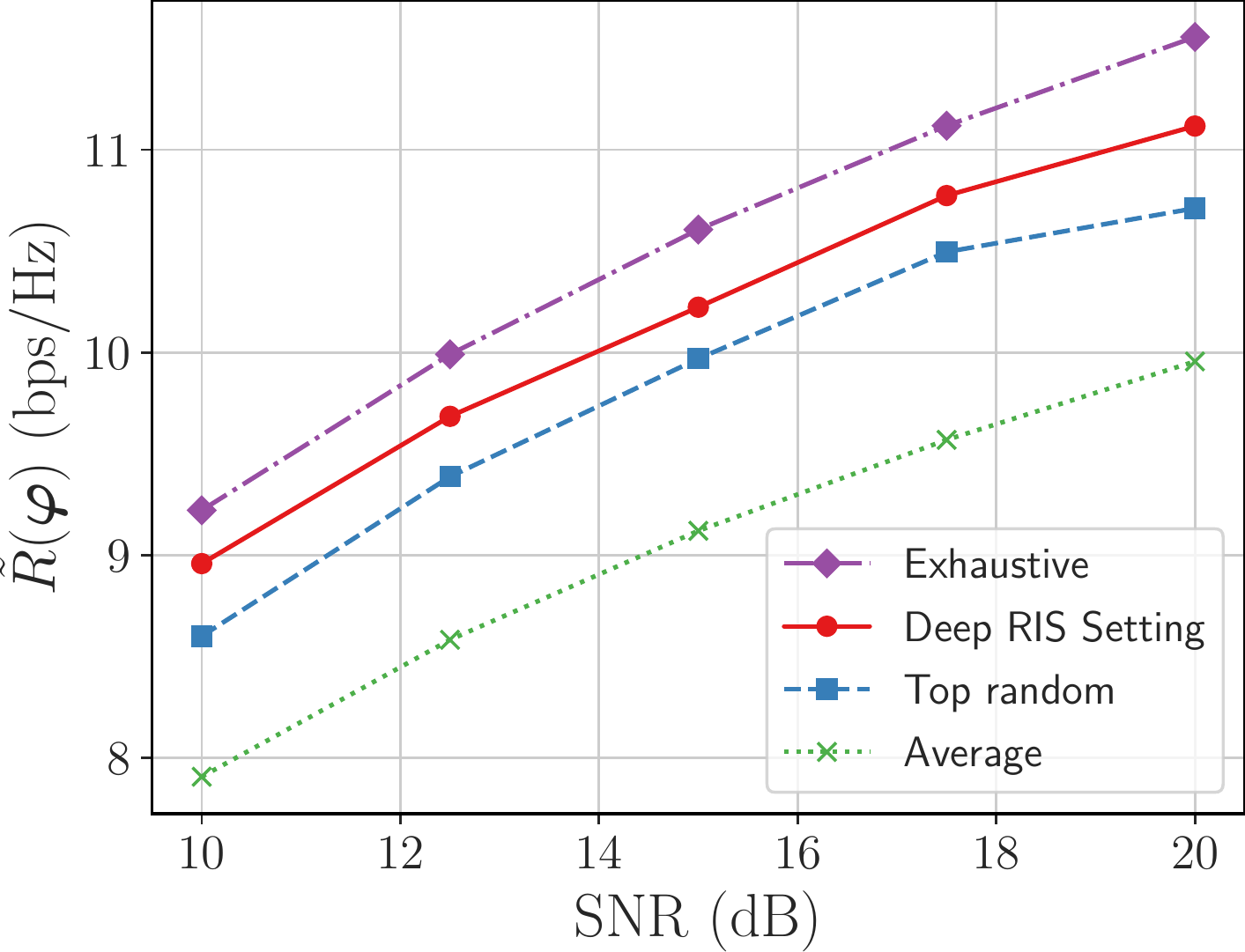}
     \vspace{-0.3cm}
    \caption{Ergodic rate performance of the proposed method and baselines over different \ac{snr} values.}
    \label{fig:results-plot}
\end{figure}

\begin{figure}[tb]
     \centering
     \includegraphics[width=0.4\textwidth]{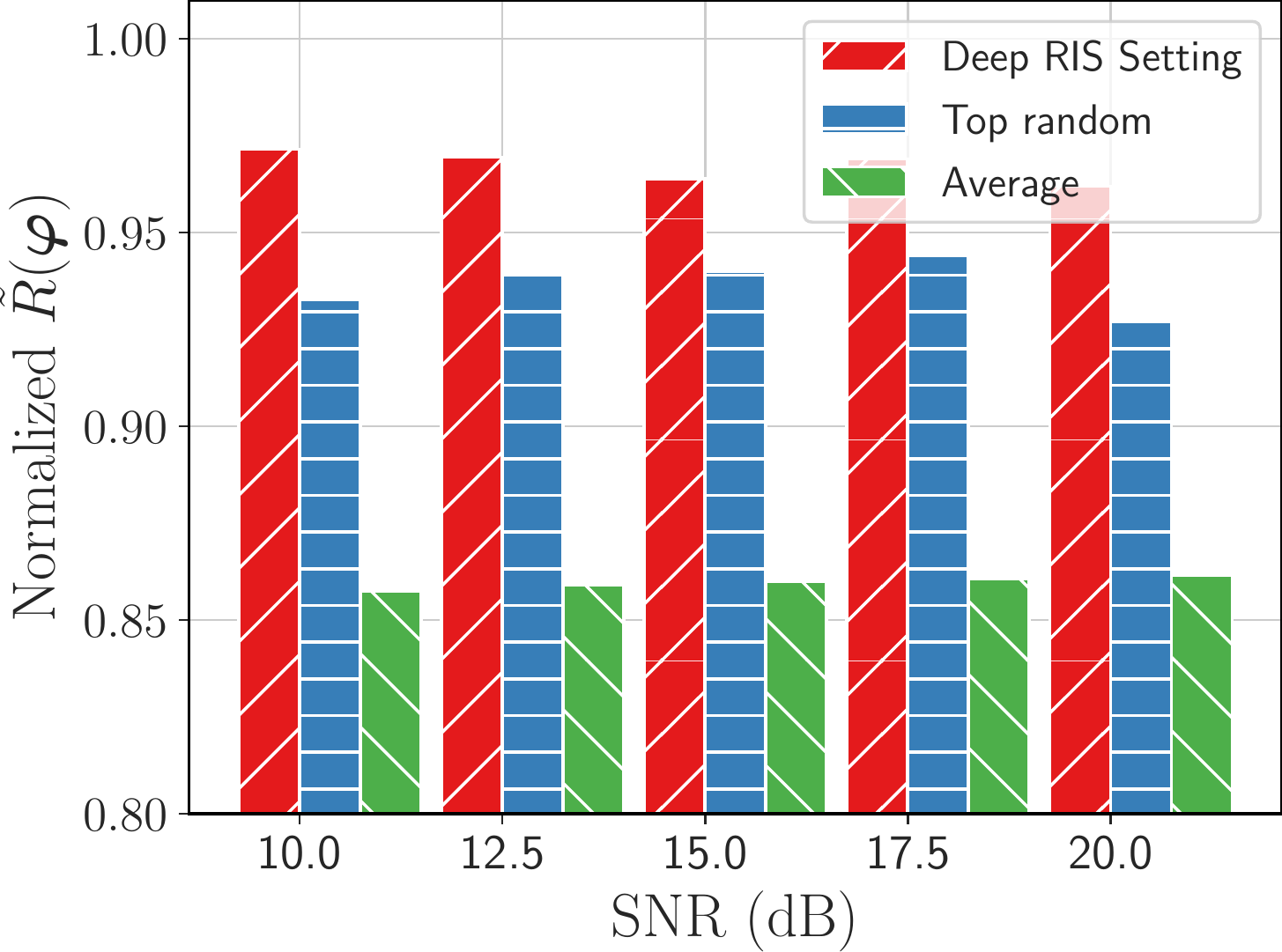}
     \vspace{-0.3cm}
    \caption{Ergodic rate of the proposed method as a fraction of the optimal achievable rates of exhaustive search.}
    \label{fig:results-bar}
\end{figure}

The achieved performances are plotted in Fig.~\ref{fig:results-plot}, whereas a normalized presentation, produced by dividing by the maximal achievable rate (via exhaustive search), is given in Fig.~\ref{fig:results-bar}.
It is shown that Algorithm~\ref{alg:deep-RIS-setting-genetic}, which adopts a \ac{dnn}-aided \ac{ga}, is capable of providing consistent improvements over the considered baselines. While the improvement offered by Algorithm~\ref{alg:deep-RIS-setting-genetic} over the Top random strategy is approximately $1$ dB in \ac{snr}, the margin for improvement in the studied scenario is limited \textit{a priori}. This is illustrated by the facts that: (i) the optimal rate is only $15\%$ greater than the random, and (ii) our proposed method performs close to the optimal performance.

\vspace{-0.2cm}
\section{Conclusion}
\label{sec:Conclusions}
\vspace{-0.1cm}
In this paper, considering dynamic rich-scattering environments, we presented an RIS configuration algorithm which bypasses the need to analytically model or experimentally measure the relationship between the \ac{ris} and channel statistics, and instead learns it from data. Based on a low-\ac{snr} approximation of the ergodic rate, this learned model is then combined with a \ac{ga} to optimize the \ac{ris} phase configuration. Our numerical study demonstrates the capability of our methodology to tune an \acp{ris} to boost reliable communications in dynamic rich-scattering environments.

\bibliographystyle{IEEEtran}
\bibliography{IEEEabrv,refs}
\end{document}